\documentclass[useAMS,usenatbib]{mn2e}
\usepackage{graphicx}

\title[A family of cometary globules]{A family of cometary globules at the periphery \protect \\ of Cyg OB1: the star HBHA 3703-01 and the reflection nebula GM 2-39}
\author[V. P. Arkhipova et al.]{V. P. Arkhipova\thanks{E-mail:
vera@sai.msu.ru}, O. V. Egorov, V. F. Esipov, N. P. Ikonnikova,\newauthor T.
A. Lozinskaya\thanks{E-mail: lozinsk36@mail.ru}, G. M. Rudnitskij, T. G.
Sitnik\thanks{E-mail: sitnik@sai.msu.ru}, A. M. Tatarnikov,\newauthor D. Yu. Tsvetkov, A. V. Zharova\\
Lomonosov Moscow State University, Sternberg Astronomical
Institute, 13 Universitetskij prospekt, Moscow 119234 Russia}
\begin{document}

\date{Accepted 2013.... Received 2013...; in original form 2012...}

\pagerange{\pageref{firstpage}--\pageref{lastpage}}
\pubyear{2013}

\maketitle

\label{firstpage}

\begin{abstract}
The interstellar medium in the region of a family of cometary
globules including the reflection nebula GM~2-39 has been analyzed
basing on our observations with the slit spectrograph, the results
of our previous observations with a Fabry--Perot interferometer in
the H$\alpha$ line, \emph{Spitzer} archival data together with CO
microwave data. The structure of globules' IR emission, velocity
field of ionized gas in the H$\alpha$ line and of molecular gas in
the CO line have been considered.  We have detected a CO cavern
around the eastern globules and faint high-velocity H$\alpha$
features of surrounding gas. The most probable sources of ionizing
radiation and wind of the Cyg OB1 association responsible for the
globules' formation are proposed. Based on our multicolour
photometry, it has been found that the brightest compact source in
the southern `head' of the cometary globule -- the star HBHA
3703-01 (IRAS 20153+3850) -- is a B(5--6)V star with
$E(B-V)$=1.18 mag. The strong and broad H$\alpha$ emission line was
revealed in its spectrum. The spectral energy distribution of
HBHA~3703-01 in the 0.44--24~$\mu$m range has been modeled. It is
shown that the star has a hot dust envelope with
$T_{\mathrm{dust}}=1400$~K and $\tau_V=1.1$. Besides, the star
HBHA 3703-01 illuminates the encompassing diffuse nebula GM 2-39
with a diameter of about 30~arcsec. With regard to all obtained
observational data for HBHA 3703-01, this object may be classified
as the Herbig Ae/Be star.
\end{abstract}

\begin{keywords}
ISM: kinematics and dynamics -- ISM: clouds -- ISM: lines and bands -- infrared: ISM -- ISM:
individual objects: GM 2-39 -- stars: individual: HBHA 3703-01
\end{keywords}

\section{Introduction}
\label{sec:intro}

The concept of a \emph{globule} was introduced as early
as in 1940s for `dark nebulae' -- opaque compact clouds
with sharply delineated boundaries on the background of
emission nebulae. Described for the first time by
\citet{haw76} and \citet{san76}, cometary globules (CGs)
are characterised by cometlike morphology -- a compact
opaque dust `head', frequently surrounded by a bright
rim, and a long dust `tail'. It is supposed that such a
shape of CGs is formed as a result of the action of
ultraviolet radiation \citep{rei83} and/or of shock
waves initiated by supernovae or stellar winds
(\citealt{bra81,bra83,sahu88,pit11}) on small dense
clouds immersed in a rarefied medium. As a rule, CGs'
tails are oriented radially from the source that has
formed them. The problem of the origin of CGs is still
discussed. The study of their nature is of special
interest because of their connection with the process of
star formation: in heads of many globules compact IR
sources -- young stellar objects (YSOs) -- have been
detected.

The launch of IR space observatories opened a new stage
in the study of CGs formation processes and their
generic connection with star-forming regions. Dense dust
clouds and their `fine structure', formed by radiation
and wind from young stars and determining the process of
formation of a new generation of stars, are studied
today very intensely using IR observations. The most
impressive results have been obtained for the Carina
Nebula (NGC~3372) from observations with the
\emph{Spitzer} and \emph{Hubble} space telescopes
\citep{sm10a,sm10b}. Here, multiple new Herbig--Haro
objects with distinct bipolar jets, cometary globules
with signs of star formation, circumstellar gas-dust
discs (proplyds) have been detected. The spatial
distribution of IR emission in different spectral bands
revealed a clear picture of continuing star formation in
dust pillars, which result from the action of young
stars on the remnants of the parent molecular cloud.

In this paper we study the structure and kinematics of
the interstellar medium towards a family of CGs
($\alpha_{2000}\sim
20^{\mathrm{h}}15^{\mathrm{m}}40^{\mathrm{s}}\textrm{--}
20^{\mathrm{h}}17^{\mathrm{m}}45^{\mathrm{s}}$,
$\delta_{2000}\sim38^{\circ}
57^{\prime}\textrm{--}39^{\circ}10^{\prime}$) detected
in the archival data of the \emph{Spitzer} Space
Observatory in the field of the supershell in Cygnus. We
carried out spectral observations of a compact stellar
object in the head of the southern globule. The purpose
of this work is the analysis of kinematics of ionized
and molecular gas, revealing a physical association
between different components of the stellar and gas
populations of the region and search for traces of the
Cyg~OB1 stars' wind action on cometary globules.

The description of IR images of CGs and their comparison
with optical images is given in Section~\ref{sec2}. In
Section~\ref{sec3} we report the results of our study of
the kinematics of the ionized gas towards the family of
CGs using the data of our long-term interferometric
H$\alpha$ observations of the gas-dust complex in
Cygnus. In Section~\ref{sec4} we present evidence for
the association of the family of IR globules with a
complex of molecular clouds and stars at the periphery
of Cyg~OB1; we have searched for probable kinematic
signs testifying to the action of stellar wind on gas
near CGs. In Section~\ref{sec5} we present the results
of photometric and spectral observations of the star
HBHA 3703-01 and the model of the dust envelope around
this star. In Section~\ref{sec6} we summarize the main
results together with their implications.

\section{Family of cometary globules in the infrared and optical ranges}
\label{sec2}

The family of cometary globules in the region $l\sim
76^{\circ}30^{\prime}$--\,$76^{\circ}40^{\prime}$;
$b\sim1^{\circ}55^{\prime}$--\,$2^{\circ}15^{\prime}$ was
detected in the infrared using observational data of the
\emph{Spitzer} Space Observatory obtained from the \emph{Spitzer}
Legacy Survey of the Cygnus-X Complex\footnote{\emph{Spitzer}
Proposal, ID 40184 (2007)} (\citealt{hora08},
\texttt{http://www.cfa.harvard.edu/cygnusX}). The archival data
were downloaded from the \emph{Spitzer} Heritage
Archive\footnote{\texttt{http://sha.ipac.caltech.edu}}. We had at
our disposal images of the region in five wavebands: 3.6~$\mu$m,
4.5~$\mu$m, 5.8~$\mu$m, 8.0~$\mu$m observed with the InfraRed
Array Camera (IRAC) and 24~$\mu$m with the Multi-Band Imaging
Photometer for \emph{Spitzer} (MIPS). The spatial resolution of
the images obtained with these cameras is 0.6 and 2.45~arcsec per
pixel, respectively. We composed mosaics of the images using the
MOPEX
software\footnote{\texttt{http://ssc.spitzer.caltech.edu/dataanalysistools/tools/mopex}}.

\begin{figure*}
\includegraphics[scale=0.8]{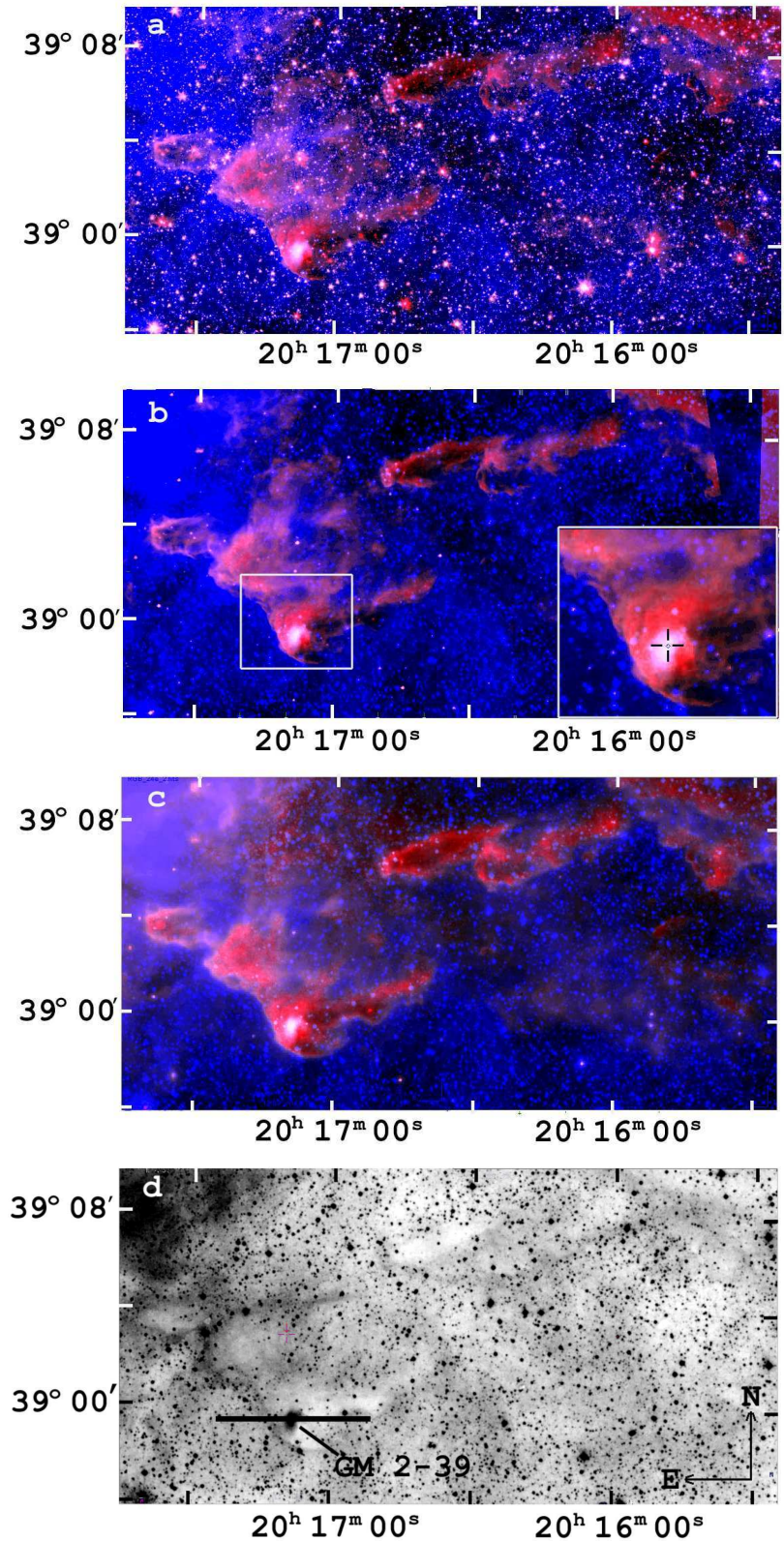}
\caption{The family of cometary globules in the bands 3.6~$\mu$m
(a), 8~$\mu$m (b), 24~$\mu$m (c) according to the \emph{Spitzer}
data (the IR image shown by red is combined with the optical
shown by blue) and in the optical range (Palomar Sky Atlas,
E-map) (d). Right lower corner of Fig.~b: enlarged map of the
southern head of the eastern globule with the embedded cometary
nebula GM 2-39; a cross marks the star HBHA 3703-01. The
localization of the slit spectrogram is indicated (Fig.~d).}
\label{fig1}
\end{figure*}

Figures~\ref{fig1}abc show the images of the region in
the 3.6, 8 and 24~$\mu$m bands. It is visible that the
family of IR globules consists of several lined in a
chain, identically oriented CGs with sizes from $\leq 1$
to 7~arcmin. The morphology of the IR globules in all
the wavebands is the same as a whole. (The `blurred'
image in the 24~$\mu$m band is due to a lower resolution
and, probably, to the fact that heated dust responsible
for the main emission in this band can be present also
in the neighbourhood of the globules.) The IR images of
the globules coincide with the regions of opaque dust in
Fig.~\ref{fig1}d surrounded with gas emitting in the
optical range. The bright contours of the IR globules
are \textbf{inside} their optical boundaries.

\begin{table*}
 \centering
 \begin{minipage}{140mm}
 \caption{Radial velocities of ionized hydrogen}
\label{table1}
\begin{tabular}{lcrcr}
  \hline
 Location & $ \emph{l,b}$ & N1&
$(\langle V_{\mathrm{LSR}}\rangle -\Delta)\ldots$ & N2 \\
 & $\alpha _{2000}, \delta_{2000}$&&$(\langle V_{\mathrm{LSR}}\rangle +\Delta)$, km~s$^{-1}$& \\
(1)&(2)&(3)&(4)&(5)\\
  \hline
   Eastern globules & $76^{\circ}32^{\prime}$--\,$76^{\circ}40^{\prime}$&36& $\mathbf{4\ldots 6}$&11 \\
      & $1^{\circ}52^{\prime}$--\,$2^{\circ}06^{\prime}$ & &$-55\ldots -35$ &12 \\
      &                                               & &$40\ldots 44$ &4 \\
& & & & \\
   Western globules & $20^{\mathrm{h}}15^{\mathrm{m}}58^{\mathrm{s}}$--\,$20^{\mathrm{h}}16^{\mathrm{m}}50^{\mathrm{s}}$ &30& $\mathbf{3\ldots 7}$ &16 \\
       & $39^{\circ}05^{\prime}$--\,$39^{\circ}09^{\prime}$& &$-56\ldots -28$ &11 \\
       &                                              & &$33\ldots41$ &3  \\
& & & & \\
   CO cavern around WR 139 & $20^{\mathrm{h}}17^{\mathrm{m}}00^{\mathrm{s}}$--\,$20^{\mathrm{h}}20^{\mathrm{m}}00^{\mathrm{s}}$ &230& $\mathbf{6\ldots 14}$&147 \\
   (region II)   & $38^{\circ}20^{\prime}$--\,$38^{\circ}45^{\prime}$& & $-60\ldots -50$&12 \\
      &                                              & & $-32\ldots-20$&28 \\
& & & & \\
   CO cavern near HD 193595 & $20^{\mathrm{h}}18^{\mathrm{m}}18^{\mathrm{s}}$--\,$20^{\mathrm{h}}19^{\mathrm{m}}20^{\mathrm{s}}$&63& $\mathbf{7\ldots 15}$&45 \\
   and HD 228841 (region III)& $38^{\circ}52^{\prime}$--\,$39^{\circ}10^{\prime}$& & $-35\ldots -31$&7 \\
   \hline
\end{tabular}
\end{minipage}
\end{table*}

In the IR range as in the optical the `three-head' structure and
the tail of the eastern IR globule are visible; the size of the
largest head reaches 3~arcmin. The heads are `filled' with IR
emission of different structure: the northern oval head is limited
with a bright contour and has a cellular IR structure
(Fig.~\ref{fig1}b); in the `middle', cone-shaped one, the
brightness of IR emission decreases toward the tail. The brightest
southern head hosts the reflection nebula GM~2-39 detected in the
optical range by \citet{gyul77} (Fig.~\ref{fig1}d). Its size does
not exceed 1~arcmin, and the integrated stellar magnitude on
Palomar Sky Survey images is nearly identical in the red and blue
light. GM~2-39 is the brightest extended IR source in the area;
its morphology in the optical and IR ranges is identical
(Fig.~\ref{fig1}b). In the western IR globules having an elongated
shape and smaller size, we reveal bright, sometimes intermittent
contours with a brightening in the region of the supposed head and
at the end of the tail. (Note that the studied objects are not
classical cometary globules, typical of which is not only a
distinct bright head, but also a long dust tail. Probably, the
western globules are fragments of the tails of the eastern heads;
a `classical' cometary nebula is only the southern part of the
eastern globule.)

Peculiarities of the globules in the IR range are thin small filaments in the
form of `whiskers' near the heads (Fig.~\ref{fig1}ab). Similar IR filaments
have been detected in a large number of cometary globules. As an example, let
us mention globules in the Carina Nebula \citep{sm02}. The filaments in front
of the southern head are curled toward the tail; they resemble rims
noticeable in the optical range at the peripheries of the heads of cometary
globules in the Gum Nebula \citep{rei83}. From the east the `Three-headed' IR
globule is also bordered with an optical rim, a much broader one than that of
IR (Fig.~\ref{fig1}d). Between the IR head and its optical rim a sharp
boundary, especially prominent in the case of the southern head, is seen
(Fig.~\ref{fig1}b). The western globules are limited with bright IR contours;
however, there are no optical counterparts for them. Note that there are
regions of cool dust in the tails of the eastern and one of the western
globules (Fig.~\ref{fig1}).

Thus, the morphology of CGs in the IR bands is as a whole
consistent with the optical images: dust head and tail, restricted
with bright IR and/or optical rims. The optical emission encircles
the IR globules. However, in the IR range there is a
characteristic feature that has no optical counterpart: thin
filaments in front of the heads of the `Three-headed' CG. The
structure of other globules in the IR range is also more
expressive than in the optical, since at the periphery the
globules are bordered with bright IR rims in all considered
spectral bands.

\section{Velocities of ionized gas in the region studied}
\label{sec3}

Detecting physical connection between different
components of ionized gas, dust, dense molecular clouds
and stellar population of the extended region around the
considered CGs requires their radial velocities. It is
difficult to find such a connection, because the region
belongs to a heavily populated field, that is viewed
along the spiral arm.

In the analysis of velocities of ionized gas towards CGs and their
neighbourhood we used the results of our observations of the
Cygnus gas-dust complex in the H$\alpha$ line carried out with a
Fabry--Perot interferometer (IFP) mounted in the Cassegrain focus
of the 125-cm reflector of the Crimean Laboratory (Sternberg
Astronomical Institute, Moscow State University). A technique of
observations and their processing were described in detail by
\citet{loz98}. The field of view and angular resolution of the
observations was 10~arcmin and 3--4~arcsec, respectively; the
actual spectral resolution was 10--15~km~s$^{-1}$. The line
profile was approximated with one or several gaussian curves in
the supposition that the FWHM of each component is larger than the
halfwidth of the instrumental profile and signal-to-noise ratio is
$\geq 5$.

In 1993--1995 we obtained 15 IFP-images in a broad
vicinity of CGs. Here, as everywhere in the Cygnus
complex, multipeak profiles of the H$\alpha$ line are
observed. They consist of one or several bright main
components with velocities of the peaks within the
interval $V_{\mathrm{LSR}}\sim
0\textrm{--}20$~km~s$^{-1}$ and of faint `shifted'
components in the line wings with velocities up to
$V_{\mathrm{LSR}}\sim -80$~km~s$^{-1}$ and
$V_{\mathrm{LSR}}\sim 60$~km~s$^{-1}$. The radial
velocities of the main component are due to systematic
motions created by the galactic rotation and spiral
density shock waves. Those of the shifted components are
caused by local high-velocity flows of gas related to
shock waves initiated by supernovae and/or stellar
winds. At large negative velocities we cannot eliminate
also a contribution from distant HII regions of the
Perseus Arm.

For allocation of characteristic systematic motions of
ionized hydrogen in the field of CG and their
neighbourhoods we have plotted `diagrams of occurrence'
of radial velocities in selected areas. Using them, we
have found radial velocities of gas $V_{\mathrm{LSR}}$
characteristic of the region as the average values at
half-maximum of the corresponding diagram together with
root-mean-square errors of their estimate $\Delta$. The
obtained velocities are listed in Table~{\ref{table1}}
in columns: (1)~region identification (CO caverns are
shown in Fig.~\ref{fig2}b); (2)~Galactic or equatorial
coordinates; (3)~number of points N1 at which radial
velocities were measured; (4)~range of the prevalent
radial velocities of systematic motions $(\langle
V_{\mathrm{LSR}}\rangle -\Delta)\ldots (\langle
V_{\mathrm{LSR}}\rangle +\Delta)$ (emphasized in
boldface); (5)~number of measurements N2 from which the
velocity range was determined. Given for each of the
four areas in the first line are velocities of
systematic motions of gas, in the second and third lines
are velocities of shifted positive and negative
components of the H$\alpha$ line. The field of radial
velocities in the broad neighbourhood of globules
considered here was analyzed also by \citet{esi96} and
\citet{sit11}.

The detected intervals of systematic radial velocities of ionized
hydrogen are typical of the population of the Cygnus Arm and are
consistent with the corresponding velocities of molecular clouds
of this arm according to the data of \citet{leu92} (see also
\citealt{sit09}) and with modern CO observations of the Cyg~X
region with a high angular resolution (\citealt{got12} and
references therein).

\section{Stellar population, structure and kinematics of gas
in the vicinity of the cometary globules}
\label{sec4}

\begin{figure*}
\includegraphics[scale=0.85]{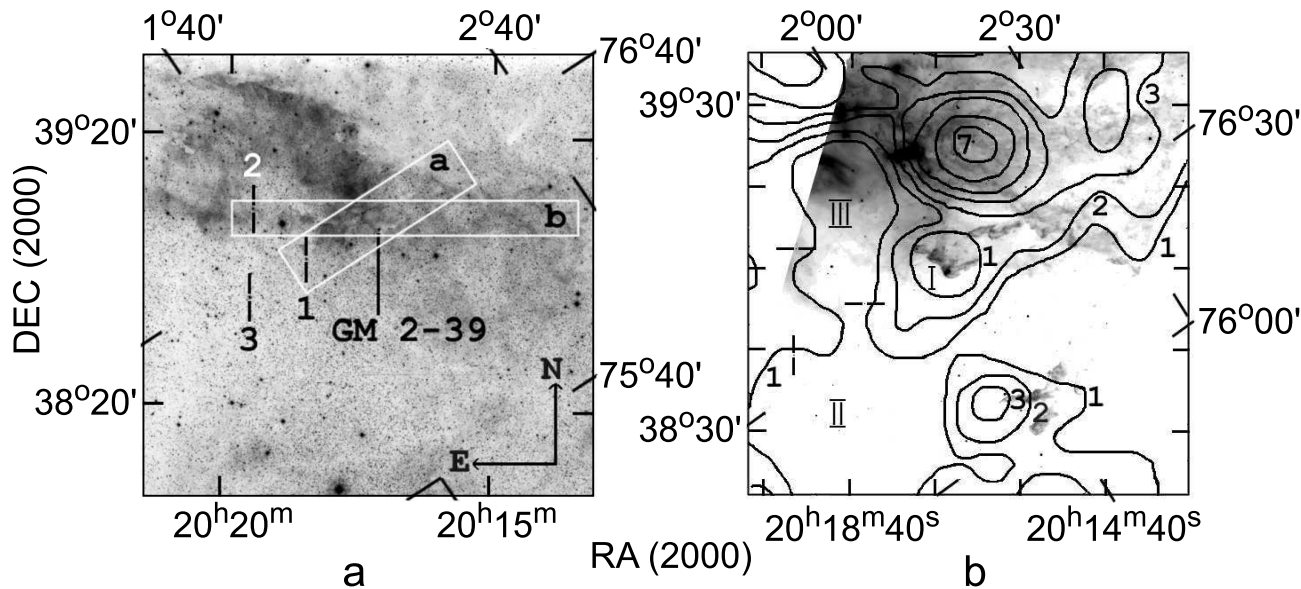}
\caption{Neighbourhood of the family of cometary globules. (a)~--
The optical map in the red range. Numbers \emph{1}, \emph{2} and
\emph{3} label the stars HD 228841, HD 193595 and WR~139.
Rectangles limit the areas for which the P/V diagrams were
plotted. The nebula Simeiz 55 is in the area ($\alpha_{2000}\sim
20^{\mathrm{h}}13^{\mathrm{m}}- 20^{\mathrm{h}}18^{\mathrm{m}}$,
$\delta_{2000}\sim38^{\circ}30^{\prime}- 39^{\circ}20^{\prime}$).
(b)~-- Image of the region in the 24~$\mu$m band with contours of
the CO emission at $V_{\mathrm{LSR}}\simeq 2.5-5$~km~s$^{-1}$,
data of \citet{leu92}. Contours are labeled in the order of
increasing brightness: 1 corresponds to the weakest CO emission.
Roman numerals: CO caverns mentioned in the text and
Table~{\ref{table1}}. Ticks mark the same stars as in Fig.~a.}
 \label{fig2}
\end{figure*}

According to adopted concepts
\citep{rei83,bra81,bra83,sahu88,pit11}, cometary globules are
shaped by shock waves and/or ultraviolet radiation, so the head
should point to the energy source, and the tail is extended
radially from it. In our case individual CGs as well as the
entire family of CGs lined in a chain are oriented such that the
heads `look' towards the plane of the Galaxy (see
Fig.~\ref{fig2}a). Accordingly, the energy source should be sought
for in a region southeast/east from CGs. Note that the considered
region includes many optical globules
(Fig.~\ref{fig1}d,~\ref{fig2}a); however, not all of them show up
in the IR range and not all are surrounded by bright optical rims.

We can reveal toward the globules and their neighbourhood in the
southeast sector of the complex the following components of the
interstellar medium and stellar population.

1.~\textbf{A CO cavern ($\mathbf{\emph{l}\sim 76^{\circ}
30^{\prime}\textrm{--}\,\,76^{\circ}
40^{\prime},\,\,\emph{b}\sim
1^{\circ}55^{\prime}\textrm{--}\,\,2^{\circ}
15^{\prime}}$) is observed at the velocity
$\mathbf{\emph{V}_{\mathbf{LSR}}\simeq
2.5\textrm{--}7.5}$~km~s$^{-1}$ \citep{leu92} towards
the eastern globules}(region I in Fig.~\ref{fig2}b).
According to Table~\ref{table1}, ionized hydrogen that
the globules are embedded in radiates at the same
systematic velocities $V_{\mathrm{LSR}}\simeq
4\textrm{--}6$~km~s$^{-1}$. In a 36~arcmin region
including the family of globules, the emission in the
H166$\alpha$ radio recombination line at velocity
$V_{\mathrm{LSR}}\simeq 5.4$~km s$^{-1}$ is also
observed \citep{lan83}.

2.~\textbf{CGs are at the periphery of the young stellar
association Cyg~OB1, inside a large-scale shell  around
Cyg~OB1 and Cyg~OB3} detected in the optical, radio and
IR ranges \citep{bra77,loz88,loz90}. The distance to
Cyg~OB1 is 1.5~kpc \citep{gar92,sit96}. The median
radial velocity of the stars of Cyg~OB1 found from the
velocities for 34 of 70 stars of the association is
4~km~s$^{-1}$ with a dispersion of 8.9~km~s$^{-1}$
\citep{sit03}, which is consistent with the quoted above
velocities of ionized gas and CO emission and allows us
to suggest a close spatial location of Cyg~OB1 and
globules. Note that in the studied direction of the
Cygnus Arm, except for the cluster Be~86 which is a
member of Cyg~OB1 \citep{for92}, no other young stellar
group has been detected.

3.~\textbf{At a separation of $\mathbf{15-30~arcmin}$ to
the east and southeast from the chain of CGs, on its
extension, at least three stars with a powerful wind are
localized:  HD 228841 (Of), HD 193595 (Of) and WR~139}
(Fig.~\ref{fig2}a). Both Of stars are members of Cyg~OB1
association \citep{bla89,gar92}, and WR~139 is its
probable member \citep{vdh01}. All three stars are
observed in a region with depleted CO emission
\citep{leu92} detected at velocities $V_{\mathrm{LSR}} >
2.5$~km~s$^{-1}$ inside the supershell around Cyg~OB1,
Cyg~OB3 (regions II, III in Fig.~\ref{fig2}b). This CO
cavern, which is probably formed by combined action of
stellar wind and radiation of the stars of Cyg~OB1
\citep{loz88,sit11}, is limited from the north by a
complex of molecular clouds at velocity
$V_{\mathrm{LSR}}\simeq 2.5\textrm{--}7.5$~km~s$^{-1}$.
The star HD 228841 is on the extension of the chain of
CGs and is separated from the nearest heads of CGs by
6~pc (see Fig.~\ref{fig2}). The star WR~139 is located
in the same direction at a projected distance of 13~pc
from the heads of CGs. The star HD 193595 is located at
9~pc to the east from the family of globules; the chain
of CGs is partially deployed toward it. The extension of
the chain of CGs at the distance of Cyg~OB1 is 9~pc, and
the sizes of the globules are not larger than 3~pc.

4.~Toward the family of globules \textbf{high-velocity
motions of ionized hydrogen} with characteristic
velocities $V_{\mathrm{LSR}}\sim -56\ldots
-30$~km~s$^{-1}$ and
$V_{\mathrm{LSR}}\sim33\textrm{--}44$~km~s$^{-1}$ are
observed (see Table~\ref{table1}). Earlier,
high-velocity motions of ionized hydrogen in this region
were attributed to the filamentary nebula Simeiz 55
\citep{esi96} (Fig.~\ref{fig2}a). In the sky plane the
nebula is a part of the supershell around the Cyg~OB1
association. The family of globules is observed on the
background of this nebula (Fig.~\ref{fig2}a). The nature
of high-velocity motions and filamentary morphology of
Simeiz 55 are not completely clear, because in the
nebula neither synchrotron radio emission, which would
testify to a supernova explosion, nor a star with a
powerful wind have been detected.

\begin{figure}
\includegraphics[scale=0.85]{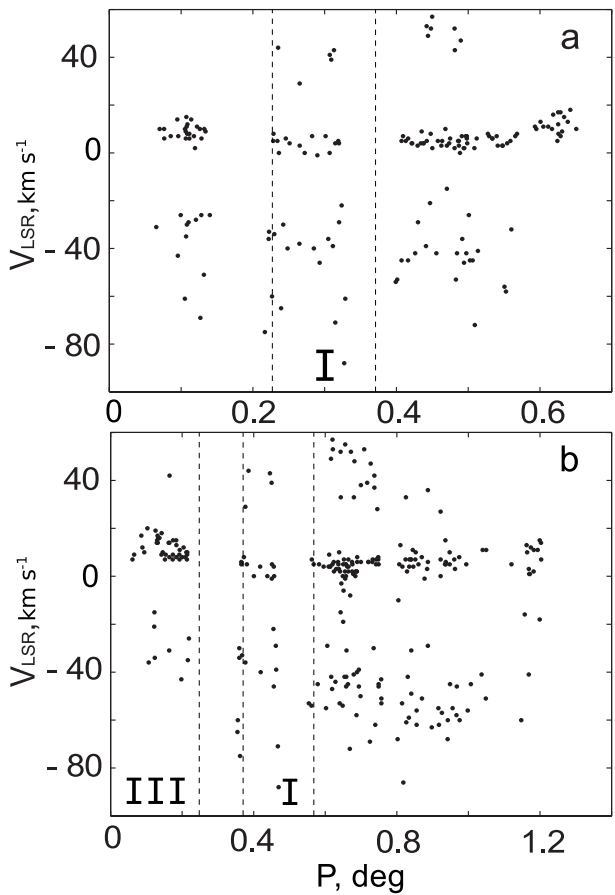}
\caption{P/V diagrams to the northwest from HD 228841 (a) and to
the west from HD 193595 (b) for the regions designated in
Fig.~\ref{fig2}a. Dashed lines denote borders of the regions I
and III shown in Fig.~\ref{fig2}b.}
\label{fig3}
\end{figure}

\textbf{High-velocity motions of ionized hydrogen} in the
neighbourhood of the family of globules could be evidence for the
physical association of cometary globules with nearby stars of the
Cyg~OB1 association. To find probable kinematic signs of the Of
stars wind action on gas in the neighbourhood of the globules, we
have analyzed high-velocity components in the H$\alpha$ line.
Along the chain of the globules, we have plotted distributions of
radial velocities as functions of projected distances to each of
the stars (so-called P/V diagrams) (Fig.~\ref{fig3}). The regions
for which P/V diagrams were plotted are shown in Fig.~\ref{fig2}a.

In Fig.~\ref{fig3}a high velocities of the shifted components in the
H$\alpha$ line (up to $-60\ldots -80$~km~s$^{-1}$) are found at a separation
of $0.\!\!^{\circ}1$ (2.5~pc) from the star HD 228841 and are visible up to
the supershell around Cyg~OB1, where they become equal to the mean velocity
of the surrounding medium. We see no definite decrease in the radial velocity
that could be attributed to a head-on collision of the shock driven by the
wind of HD 228841 with the heads of the globules. Indirect evidence for the
interaction between the wind of the star and the family of CGs can be the
shape of the heads prolate perpendicularly to the direction toward the star
as well as the presence of the CO cavern surrounding the eastern globules
(see item 1) and of fine IR filaments curled away from the star. The
existence of high-velocity motions observed beyond $0.\!\!^{\circ}25$ (7~pc),
that is farther than the heads, can be explained by gas circumflowing the
family of CGs. The main non-shifted component of the line in Fig.~\ref{fig3}
sums up the emission of non-perturbed gas near the star and in the line of
sight.

In Fig.~\ref{fig3}b high velocities, up to 60--80~km~s$^{-1}$ ,
are visible at a projected separation of $0.\!\!^{\circ}45$
(12~pc) from HD~193595, and decrease by the absolute value to the
velocity of the surrounding medium at $0.\!\!^{\circ}8$ (21~pc).
The family of globules is within the limits of this distance. Note
also that HD 193595 and the family of CGs are at the opposite
sides of the CO cavern detected at velocity $V_{\mathrm{LSR}} >
2.5$~km~s$^{-1}$ (region III in Fig.~\ref{fig2}b). In
Fig.~\ref{fig3}b this rarefied region of the interstellar medium
extends to not less than $0.\!\!^{\circ}25$ (7~pc) from HD 193595.
Therefore, high velocities of ionized hydrogen are found only far
from the star.

Thus, the family of CGs falls within the zone of action of the
wind from the Of stars HD~228841 and HD~193595. The surrounding
ionized gas swept by the wind of these stars expands at a velocity
of up to 60--80~km~s$^{-1}$ and circumflows the family of cometary
globules, since we see high velocities along the entire chain of
CGs. We have found no obvious kinematic signs of braking of
high-velocity gas due to interaction with the Three-headed
globule.

From what was said it follows that the generic association between the family
of CGs and the above-mentioned stars of Cyg~OB1 is possible. This is based on
the following arguments. The family of CGs is surrounded by ionized gas
radiating at systematic velocities $V_{\mathrm{LSR}}\simeq
3\textrm{--}7$~km~s$^{-1}$. The median radial velocity of the Cyg~OB1 stars
is 4~km~s$^{-1}$. At the same velocities a CO cavern is observed around the
eastern globules (region I in Fig.~\ref{fig2}b). This small cavern was formed
in a complex of clouds limiting another CO cavern detected at
$V_{\mathrm{LSR}}
> 2.5$~km~s$^{-1}$ and connected with the Cyg~OB1 association (see
regions II and III in Fig.~\ref{fig2}b and in Table~\ref{table1}). The
above-mentioned Of and WR stars belong to Cyg~OB1. In the sky plane the stars
lie on the extension of the chain of CGs within the limits of the action of
their UV emission and wind. Important signs of the probable physical
association of these stars with the cometary globules are radial elongation
of the chain of CGs relative to the stars and its morphology: the heads with
a bright rim point toward the stars, large bright globules are closer to the
stars than small ones. IR rims observed near the heads of globules and curled
toward the tails belong to type~1 in the classification of \citet{sm02} (see
Fig.~33 therein). This is one more argument for the association of the stars
and globules, since such a shape is formed under the action of wind and
radiation of a massive star or cluster.

\section{The star HBHA 3703-01 and the nebula GM 2-39 associated with cometary globules}
\label{sec5}

The nebula GM 2-39 hosts the star GSC 03151-00990 ($\alpha =
20^{\mathrm{h}}17^{\mathrm{m}}08^{\mathrm{s}}\!\!.05$; $\delta =
38^{\circ}59^{\prime}29^{\prime\prime}\!\!.4$ (2000)) identified with the
infrared source IRAS~20153+3850 (Fig. 1bd). The star is also included in the
catalog of H$\alpha$ emission objects \citep{koh99} as HBHA 3703-01.
According to the classification of emission spectra of stars in this catalog,
the H$\alpha$ emission line in the spectrum of HBHA 3703-01 has a medium
intensity on the background of a moderate continuum. The catalog lists the
star' brightness of $V = 13.6$~mag.

To get more complete information about the star and surrounding
interstellar medium we carried out the spectral and photometric
observations.

\subsection{Photometric observations of HBHA 3703-01}

The photometric observations were carried out with the 60-cm
reflector of the Crimean Observatory of the Sternberg Astronomical
Institute equipped with Apogee AP-47p camera.  A total of
158~images in the $UBVR_{c}RI$ bands were obtained in
September--November
2012. All reductions 
were made using IRAF\footnote{IRAF is distributed by the National
Optical Astronomy Observatory, which is operated by AURA under
cooperative agreement with the National Science Foundation}, MAXIM
and programs of V.P.~Goranskij.

\begin{figure}
\centering
\includegraphics[scale=0.5]{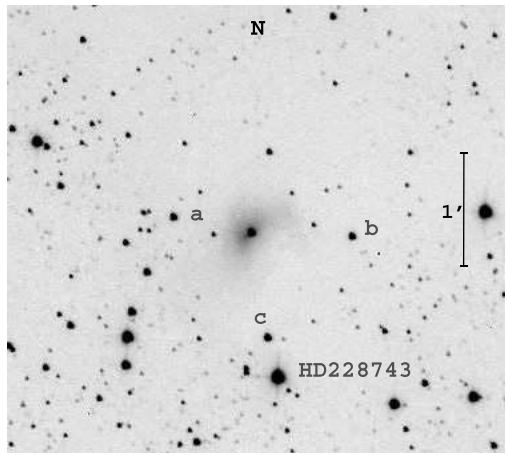}
\caption{The image in the $R$ band of HBHA 3703-01 with comparison
stars.}
\label{fig4}
\end{figure}

The comparison stars were calibrated with respect to the star HD
228743. Its stellar magnitudes ($U=11.71$~mag, $B=11.51$~mag,
$V=11.22$~mag, $R_{c}=11.04$~mag, $R=10.93$~mag and $I=10.72$~mag)
were obtained using the following photometric standards: PG2336,
PG0231, RU149 \citep{land92}, M92, NGC~7790 \citep{stet00}, M81
\citep{rich94}. The reduction of $R_{c}$ to $R$ and $I_{c}$ to
$I$ was done with equations from \citet{bes79}. The image of HBHA
3703-01 and comparison stars are shown in Fig.~\ref{fig4}.

The colour terms for transformation of instrumental magnitudes
$ubvr_{c}i$ to standard $UBVR_{c}I$ were determined  using
observations of standards in M~92 and NGC~7790 \citep{stet00} and
in M82 \citep{rich94}. The equations $u = U + K_{u}(U - B) +
C_{u}$; $b = B + K_{b}(B - V) + C_{b}$; $v = V + K_{v}(B - V) +
C_{v}$; $r_{c} = R_{c} + K_{rc}(V - R)_{c} + C_{rc}$ and $i = I +
K_{i}(R_{c}-I) + C_{i}$ were solved for colour terms:
$K_{u}=-0.08$, $K_{b}=0.04$, $K_{v}=-0.03$, $K_{rc}=-0.04$
$K_{i}=0.027$.

The results of our photometric observations of HBHA 3703-01 are
presented in Table~{\ref{tab2}}. The star brightness varies from
night to night, but these variations do not exceed 3$\sigma$;
therefore, the conclusion on star photometric variability would be
premature.

\begin{table*}
\caption{$UBVR_{c}RI$ magnitudes of HBHA 3703-01 from observations in
September--November 2012} \label{tab2}
\begin{tabular}{ccccccccccccc}
\hline
JD-2400000&$U$&$\sigma_{U}$&$B$&$\sigma_{B}$&$V$&$\sigma_{V}$&$R_{c}$&
$\sigma_{R_{c}}$&$R$&$\sigma_{R}$&$I$&$\sigma_{I}$\\
\hline

56187.3&--&--&15.89&0.05&14.96&0.01&--&--&--&--&--&--\\
56216.2&--&--&15.99&0.04&14.99&0.01&--&--&13.89&0.01&13.22&0.01\\
56218.2&--&--&15.95&0.06&14.89&0.02&--&--&--&--&--&--\\
56219.2&--&--&16.02&0.03&15.01&0.02&--&--&13.86&0.01&13.26&0.02\\
56240.2&--&--&16.02&0.02&14.94&0.01&14.29&0.01&13.87&0.01&13.29&0.02\\
56243.2&--&--&16.06&0.02&15.03&0.03&14.29&0.01&--&--&13.27&0.02\\
56247.2&--&--&15.97&0.02&15.00&0.02&14.28&0.01&--&--&13.17&0.02\\
56252.1&16.45&0.08&16.04&0.03&15.02&0.05&14.27&0.01&--&--&13.30&0.02\\
\hline

mean&16.45&0.08&15.99&0.10&14.98&0.07&14.28&0.02&13.87&0.02&13.25&0.04\\
\hline
\end{tabular}
\end{table*}

\begin{figure}
\centering
\includegraphics[scale=0.2]{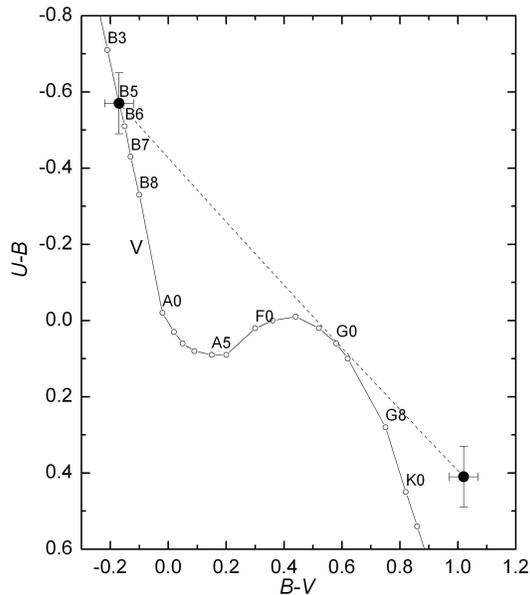}
\caption{HBHA 3703-01 in the $U-B$, $B-V$ diagram. The solid curve
is the main sequence according to \citet{str77}; the dashed line
represents the reddening line in form
$E(U-B)/E(B-V)=0.796+0.020E(B-V)$.} \label{fig5}
\end{figure}

The comparison of mean colour indices of HBHA 3703-01,
$U-B=0.4\pm0.14$~mag, $B-V=1.01\pm0.13$~mag, with normal
colour indices of main-sequence stars shows that HBHA
3703-01 is a B(5--6)V star with the colour excess
$E(B-V)=1.18\pm 0.17$~mag. For that we used intrinsic
colour indices from \citet{str77} and the reddening line
of Cyg OB2 in form $E(U-B)/E(B-V)=0.796+0.020E(B-V)$,
derived by \citet{tur89}. We note that there are only three stars HD 193032 (B1 III), HD 192422 (B0.5I) and HD 228841 (O7.5I) with known $UBV$-data and the
two-dimensional spectral classification in a close ($r\leq 30'$)
vicinity of HBHA 3703-01. Photometry of these stars provides an average
$E(U-B)/E(B-V) = 0.81\pm0.01$ that is in a good agreement with Turner's
data. The position of HBHA 3703-01
on the $U-B$, $B-V$ diagram is indicated in
Fig.~\ref{fig5}.

We estimated the visual absolute magnitude of HBHA 3703-01 using
our data -- $\overline{V}=14.98\pm0.07$~mag and
$E(B-V)=1.18\pm0.17$~mag -- assuming that the star belongs to the
Cyg~OB1 complex at the distance 1.5~kpc. It turned out to be
$M_{V}= 0.4\pm 0.1$~mag, in a good agreement with the visual
luminosity of B6 stars on the zero-age main sequence (ZAMS, $M_{V}=0.3$~mag,  \citealt{str82}).

The colour excess, $E(B-V)$=1.18~mag, derived by us, characterises the total light
extinction involving the interstellar and circumstellar
components. The problem of light extinction toward HBHA 3703-01 is
complicated, since the star is embedded in an extremely
inhomogeneous dust medium. The nearest to HBHA 3703-01 star with
the known reddening -- HD~228841 (Of7.5p) with $E(B-V)=0.87$~mag,
member of Cyg~OB1 -- is located at about 18~arcmin from the
emission star outside the region of large extinction. Adopting for
HBHA 3703-01 this value as its interstellar extinction, the
circumstellar reddening component of HBHA 3703-01 may be estimated
as about 0.3~mag.

The reflection nebula GM~2-39 surrounding the star HBHA 3703-01 is
well visible on images in all bands. The measured colour indices
of the nebula, $B-V=0.5\pm 0.2$~mag, $V-R=0.7\pm0.2$~mag, are
considerably bluer than that of the star: $B-V=1.01\pm 0.13$~mag,
$V-R=1.11\pm0.11$~mag, due to the strong scattering of the stellar
light in this nebula.

We note the star  HBHA 3703-01 was too much brighter in
\citet{dro07} ($V=13.44\pm0.12$~mag) than in our observations.
\citet{dro07} made the measurements with a 30$^{\prime\prime}$
aperture so their data refer to the total star's and a
considerable part of reflection nebula GM~2-39 radiation.

\subsection{Spectral observations of HBHA 3703-01 and of its environment}

Long-slit spectra at low spectral resolution were obtained with a
large-aperture spectrograph in the Cassegrain focus of the 125-cm reflector
of the Crimean Laboratory of the Sternberg Astronomical Institute in 2011 and
2012 years. We used a diffraction spectrograph with a 600~lines/mm grating.
The observations were conducted with a slit about 8~arcmin long and 4~arcsec
wide; the scale along the slit was 1.98~arcsec per pixel. As a detector an
ST-402 camera was used (array size 765$\times$510 pixels, pixel size
9$\times$9~$\mu$m). The actual spectral resolution was $\mathrm{FWHM}$ =
6.0~\AA. The typical achieved signal-to-noise
ratio ($S/N$) is about 10. The spectrograph slit was directed east--west (the localization of
the spectrograms is shown in Fig.~1d). Table~{\ref{tab3}} gives the log of
observations: for each spectrum it lists the date, spectral range $\Delta
\lambda$ and total exposure $T_{\mathrm{exp}}$. The spectrophotometric
standards 40~Cyg and 57~Cyg \citep{glu98} were used for flux calibration. The
spectra were processed by the standard CCDOPS software as well as by SPE code
created by S.G.~Sergeev in the Crimean Astrophysical Observatory
\citep{ser93}. The primary processing included dark current subtraction,
extracting individual spectra from CCD images and their wavelength
calibration.

\begin{table}
\caption{Log of spectral observations} \label{tab3}
\begin{tabular}{cccc}
\hline
Date&JD-2400000&$\Delta \lambda$, \AA&$T_{\mathrm{exp}}$, s\\
\hline
05/06.10.11&55840&5700--7200&600\\
26/27.10.11&55861&5700--7200&900\\
15/16.09.12&56186&4000--5700&1800\\
10/11.10.12&56211&4000--5700, 5700--7200&1800, 1800\\
15/16.10.12&56216&4000--5700, 5700--7200&1800, 1800\\
\hline
\end{tabular}
\end{table}

The star HBHA 3703-01 was placed at the slit centre. The only line
detected in the observed spectrum of the star is broad H$\alpha$
($\mathrm{FWHM}=9.5\pm 0.5$~\AA). After excluding the instrumental
profile, $\mathrm{FWHM}$ of the line is $7.3\pm 0.5$~\AA; this
corresponds to a high velocity, $\Delta v=169\pm 25$~km~s$^{-1}$.
The large linewidth can be caused by stellar wind, star rotation
or turbulent motions in the outer envelope of the star. We
measured the equivalent widths of H$\alpha$ on four spectrograms
obtained in 2011 and 2012 and did not find any variations
exceeding the measurement errors. The mean equivalent width is
$W(\mathrm{H}\alpha)=103\pm2$~\AA. Most likely, the emission line
is formed by the collisional excitation mechanism. By this, the
object is similar to Herbig Ae/Be stars, characteristic of
star-forming regions.

We measured also the absolute flux in $\mathrm{H}\alpha$ for HBHA
3703-01 $F(\mathrm{H}\alpha)= (5.1\pm 0.5)\times
10^{-13}$~erg~cm$^{-2}$s$^{-1}$. When corrected for light
extinction with $A_{V}=3.1E(B-V)$=3.65~mag, we find
$F_{0}(\mathrm{H}\alpha) = 7.0\times
10^{-12}$~erg~cm$^{-2}$s$^{-1}$. Adopting distance to the star $d$
= 1.5~kpc found for the Cyg~OB1 association, we obtain the
luminosity in the line $L(\mathrm{H}\alpha)=4\pi
d^{2}F_{0}(\mathrm{H}\alpha)=1.8\times 10^{33}$~erg~s$^{-1}$, or
$5.9\times 10^{44}$~photons per second. A typical B(5--6)V star
delivers much less photons in the range $\lambda < 912$~\AA;
therefore, the H$\alpha$ line can not be produced by a process of
photoionization followed by recombination. As will be shown below,
HBHA 3703-01 may be classified as the Herbig Ae/Be star. The
origin of the emission lines in such stars is still under
discussion.

\begin{figure*}
\hspace{-6mm}\includegraphics[scale=0.32]{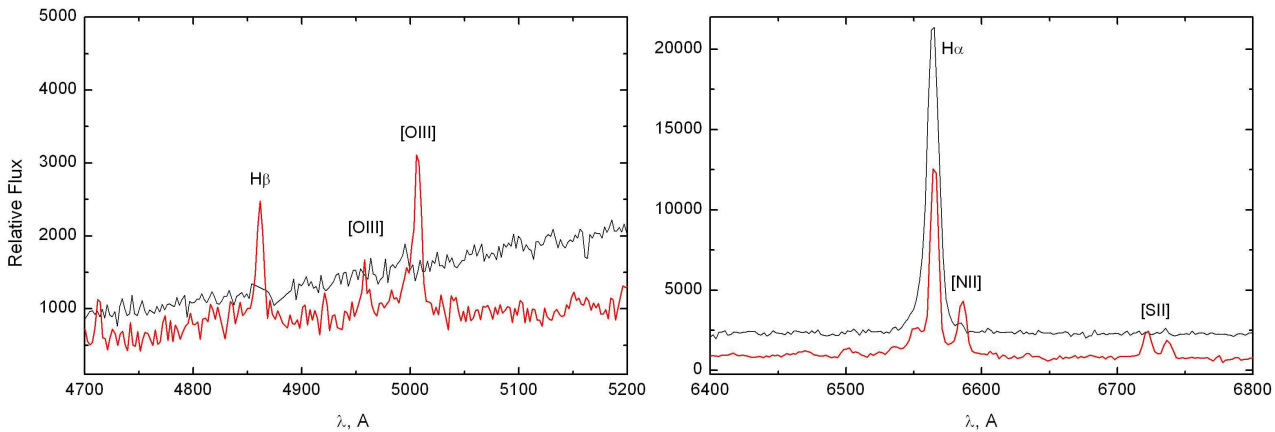} \caption{Spectrum
of the star HBHA 3703-01 (black line) and emission spectrum of
ionized gas in the region of the filament bordering the southern
head (red line).} \label{fig6}
\end{figure*}

In the spectrum of ionized gas, along the entire spectrograph slit
($\sim$8~arcmin), the emission lines H$\beta$, [O\,{\sc iii}]
$\lambda$~4959, 5007~\AA, H$\alpha$, [N\,{\sc ii}] $\lambda$~6548,
6584~\AA, [S\,{\sc ii}] $\lambda$~6717, 6731~\AA\ are seen.
Figure~\ref{fig6} shows the spectrum of the star HBHA 3703-01
after subtraction of the nebula background and the emission
spectrum of ionized gas.

The reflection nebula GM~2-39 has a diameter of about 30~arcsec
and is well seen in the optical and infrared, but it is invisible
on our spectrograms owing to its low surface brightness.

\begin{figure}
\hspace{-6mm}\includegraphics[scale=0.17]{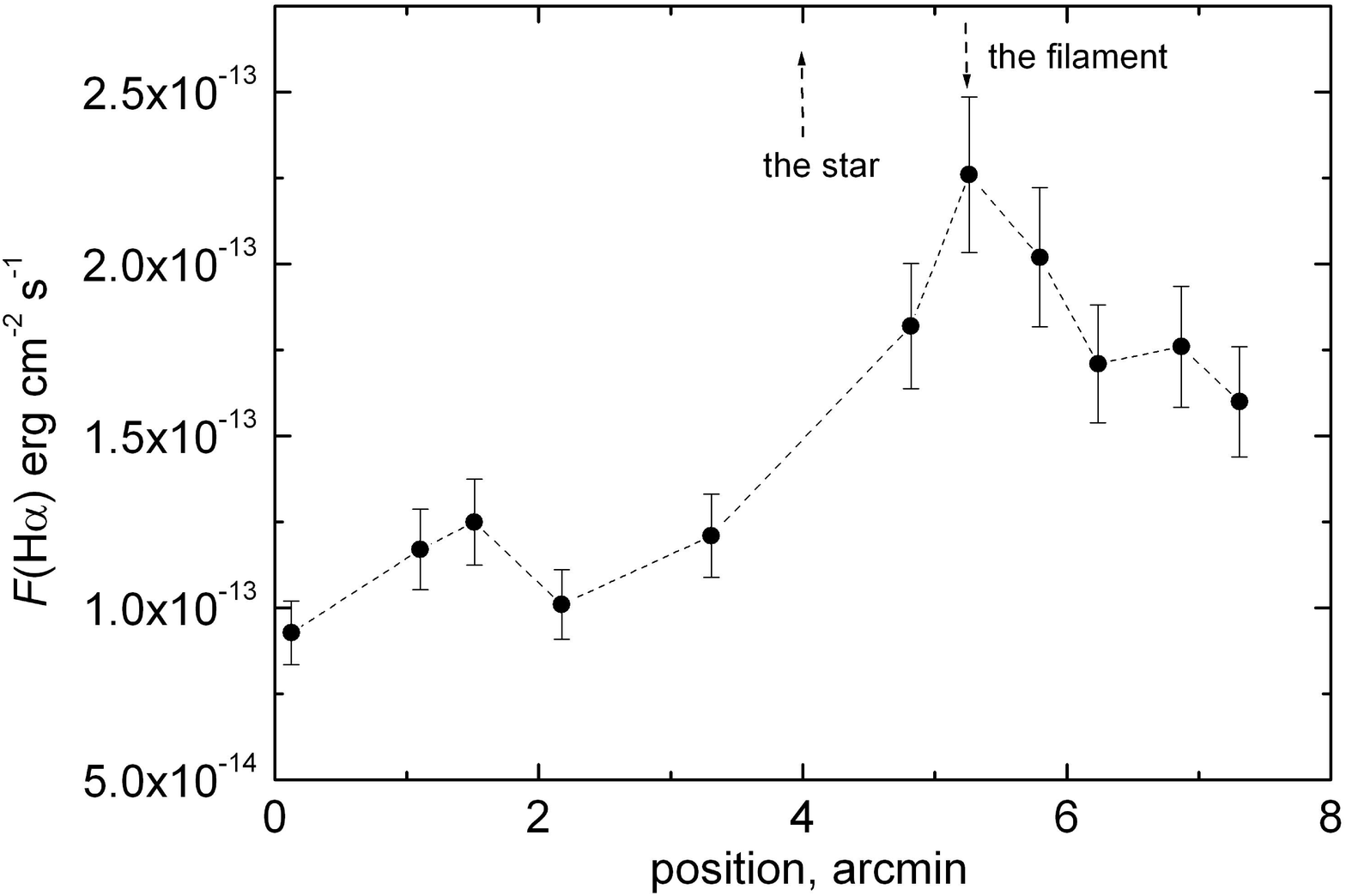} \caption{Observed
H$\alpha$ flux along the slit from west to east.} \label{fig7}
\end{figure}

The emission spectrum of ionized gas of the surroundings is due by
its origin to the general field of ultraviolet radiation of the
hot stars WR~139, HD~193595 and HD~228841. The intensity of the
emissions vary along the slit with the inhomogeneity of
interstellar medium. Figure~\ref{fig7} shows the distribution of
the H$\alpha$ flux of ionized gas along the slit from east to
west. At 1.3~arcmin to the east from the star HBHA 3703-01 there
is a dense optical filament = rim of the southern head of the
globule. On the slit at this place the emission lines of hydrogen,
sulfur and nitrogen are especially enhanced.

We measured the intensity ratio of basic emission lines of ambient
ionized gas. The average values along the slit are:
$F(5007)/F(\mathrm{H}\beta) = 1.15\pm 0.17$,
$F(6584)/F(\mathrm{H}\alpha) = 0.31\pm 0.04$,
$F(6717+6731)/F(\mathrm{H}\alpha) = 0.23\pm 0.02$. These values
are typical of the interstellar medium with solar abundances and
they fall into the position of typical H\,{\sc ii} regions on diagnostic
diagrams $\log($[N\,{\sc ii}]/H$\alpha)$ --
$\log($[O\,{\sc iii}]/H$\beta)$ and
$\log($[S\,{\sc ii}]/H$\alpha)$ --
$\log($[O\,{\sc iii}]/H$\beta)$ \citep{Kni08}.

The mean value of $F(\lambda6717)/(\lambda6731) = 1.47\pm 0.10$ is
very close to its limiting value for low densities, allowing to
determine only an upper limit of the electron density in the
ambient gas, $N_{e}\leq 40$~cm$^{-3}$.

\subsection{Model of the spectral energy distribution of HBHA 3703-01 with a dust envelope}

The object HBHA 3703-01 is well visible on original \emph{Spitzer}
images in the 3.6, 4.5, 5.6, 8, 24 and 70~$\mu$m IR bands. On all
images except for 70~$\mu$m it looks as a starlike object
surrounded by a nebula. On the last image only the nebula is
visible. Using these images except for 70~$\mu$m, we estimated
its fluxes in each of the bands in units of
$10^{-16}$~erg~cm$^{-2}$s$^{-1}$\AA$^{-1}$: $F_{3.6} = 39.2$,
$F_{4.5} = 28.8$, $F_{5.6} = 29.1$, $F_{8} = 17.1$ and $F_{24} =
18.4$. We subtracted  the nebular emission surrounding the object
from the measured fluxes. The object HBHA 3703-01 was observed
also during the 2MASS survey (its catalogue ID is
20170810+3859295); its $JHK_{s}$ magnitudes are, respectively, $J
= 11.476\pm 0.023$, $H = 10.458\pm 0.019$, $K_{s} = 9.447\pm
0.017$. The large colour index, $J-K_{s} = 2.03$, tells that the
star has an excess of IR emission most likely associated with a
hot ($>$1000~K) circumstellar gas-dust envelope.

In addition, observations of HBHA 3703-01 were obtained
by the \emph{AKARI} infrared satellite \citep{kaw07}.
According to these data, the fluxes from HBHA~3703-01 in
the 65, 90, 140 and 160-$\mu$m bands (in units of
$10^{-16}$~erg~cm$^{-2}$s$^{-1}$\AA$^{-1}$) are $F_{65}
= 68.6$, $F_{90} = 37.2$, $F_{140} = 27.0$ and $F_{160}
= 18.3$. Integration of SED in the 65--160-$\mu$m range
indicates that the total emission of dust in this range
exceeds the luminosity of the central star by a factor
of a few. Such a large flux may be due to the low
angular resolution of \emph{AKARI} (1--2~arcmin)
\citep{doi12}. In addition to the circumstellar
envelope, its field of view covers an appreciable part
of the dust cloud visible on the 24-$\mu$m image and
heated by the radiation field of the Cyg~OB1 stars.
Thus, the \emph{AKARI} fluxes refer not only to the
emission of the circumstellar envelope of HBHA~3703-01
but also to the nebula GM~2-39 and to some part of its
environment. In what follows we do not consider these
fluxes.

\begin{figure}
\includegraphics[scale=0.48]{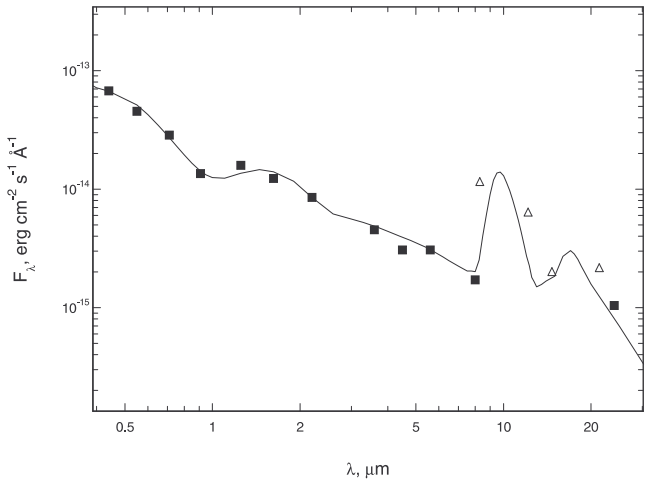}
\caption{The dereddened SED of HBHA 3703-01 in the 0.44--24~$\mu$m range
(squares), \emph{MSX} data (triangle)  and calculated energy
distribution (solid curve).} \label{fig8}
\end{figure}

Figure~\ref{fig8} shows the SED of HBHA 3703-01 in the
0.44--24-$\mu$m spectral range plotted together with our
photometric data and 2MASS ($J$, $H$, $K_{s}$), as well as fluxes
from the \emph{Spitzer} images (3.5--24~$\mu$m). The figure also
shows the 8--21-$\mu$m \emph{MSX} mission data \citep{egan03}.
All fluxes were corrected for interstellar extinction that
corresponds to $E(B-V)=0.87$~mag using the extinction law from
\citet{rie85}. This value is less than the one used above in the
estimation of the spectral type of the star, since some part of
the extinction toward the central source may be due to the
circumstellar envelope. The SED of HBHA 3703-01 taking into
account the circumstellar dust envelope was calculated using the
CSDUST3 software, which was described in detail by
\citet{egan88}. We have done the following assumptions:

1. The dust envelope was considered to be spherically symmetric
with inner and outer radii $R_{\mathrm{in}}$ and
$R_{\mathrm{out}}$, respectively. The dust density distribution in
the envelope was set as $n(r)\propto r^{-\alpha}$.

2. The spectral energy distribution of the B(5--6)V central star
was assumed to be blackbody with temperature $T_{\mathrm{eff}}$ =
15,000 K and luminosity $L_{*} = 110 \textrm{L}_{\odot}$.
Precisely this luminosity is obtained with a distance of 1.5~kpc
and with the observed bolometric flux.

3. The dust envelope consists of amorphous silicate and
graphite particles with radius $a_{\mathrm{sil}}$,
$a_{\mathrm{car}}$ and number densities
$n_{\mathrm{sil}}$ and $n_{\mathrm{car}}$, respectively.
The factors of efficiency of absorption and scattering
were calculated in accordance with the theory of
\citet{mie1908} for spherical particles. The refractive
index of the material of the particles was taken from
\citet{dor95}.  In the calculations we determined the
optical depth of the envelope at a wavelength of
0.55~$\mu$m ($\tau_{V}$), $R_{\mathrm{in}}$,
$R_{\mathrm{out}}$, $a_{\mathrm{sil}}$,
$a_{\mathrm{car}}$, $n_{\mathrm{sil}}$,
$n_{\mathrm{car}}$ and the exponent $\alpha$ in the dust
density distribution law. Figure~\ref{fig8} presents the
calculated SED of the object HBHA~3703-01, which was
obtained under the above assumptions.

The best agreement between the computed and observed SED
is achieved for the following parameters: $\tau_{V} =
1.1$, $R_{\mathrm{in}}=480~\textrm{R}_{*}$,
$a_{\mathrm{sil}}= 0.1$~$\mu$m, $a_{\mathrm{car}} =
0.25$~$\mu$m, $n_{\mathrm{sil}}$/$n_{\mathrm{car}} =
32$, $n(r)\propto 1/r^{1.5}$, $T_{\mathrm{in}} =
1350$~K.

Thus, in the circumstellar dust envelope of HBHA~3703-01
silicate particles prevail. The silicate dust
temperature at the inner boundary of the envelope is
$T_{\mathrm{in}}$=1350~K. Exponent $\alpha = 1.5$ in the
dust density distribution law is a typical value in
accreting envelopes of Herbig Ae/Be stars (see, e.g.,
\citealt{shu87}). The total mass of the dust envelope
radiating in the considered spectral range is
$M_{\mathrm{dust}}\sim 3\times
10^{-6}\textrm{M}_{\odot}$. For the ratio
$M_{\mathrm{gas}}/M_{\mathrm{dust}}\approx 100$ this
yields the total mass of the circumstellar envelope
$M_{\mathrm{env}}\sim 3\times
10^{-4}\textrm{M}_{\odot}$. Note that the estimated mass
of the envelope depends on the outer radius of the
envelope; finding the envelope mass requires
far-infrared observations with a high angular
resolution. Therefore, we estimate only the mass of the
material radiating in the wavelength range $\lambda <
25$~$\mu$m.

Thus, the object HBHA 3703-01 located at the center of the diffuse
nebula GM~2-39, in the region of the southern head of the IR
globule, is a B(5--6)Ve star with a dust envelope, whose
temperature at the inner boundary is about 1400~K. The star
illuminates the nebula GM~2-39. The nebula GM~2-39 is
distinguished from the ambient background of the molecular cloud
just because of the presence of the above-mentioned star.

Thus, the star HBHA~3703-01 satisfies completely the criteria for
Herbig Ae/Be stars \citep{Her60}: (1)~spectral type of A or
earlier with emission lines, (2)~location in an `obscured
region' and (3)~the star would illuminate a bright reflection
nebulosity in its vicinity.

In the area studied, no strong centimetre-wavelength radio
continuum source is present that could be associated with a
compact ionized region. Only one compact radio source has been
detected there, MITG J201703+3900 \citep{gri91} with flux density
$S_\nu = 53$~mJy at $\nu=5$~GHz. Most probably, it is a nonthermal
extragalactic source, its spectral index $\alpha$
($S_\nu\propto\nu^\alpha$) in the 1.4--5~GHz interval is $-$2.32.

\section{Conclusion}
\label{sec6}

Our study of the interstellar medium toward a family of cometary
globules detected in the archival data of the \emph{Spitzer}
Space Observatory in the region of a supershell in Cygnus allow
us to draw the following conclusions.

The analysis of kinematics of the interstellar medium in the
neighbourhood of the family of CGs and of the stellar population
in the regions suggests a physical association between CGs and the
Cyg~OB1 association at a distance of 1.5~kpc. (Ionized hydrogen
and CO molecular clouds near the family of CGs have
velocity $V_{\mathrm{LSR}}\simeq 3\textrm{--}7$~km~s$^{-1}$; the
median radial velocity of the stars of Cyg~OB1 is
$V_{\mathrm{LSR}}\simeq 4$~km~s$^{-1}$.)

The considered family of globules can be a remnant of a
pillar directed toward the centre of the Cyg~OB1
association. The pillar is decaying in the process of
star formation under the action of ionizing radiation
and wind of the stars of Cyg~OB1.

The family of CGs is within the zone covered by the wind and
ionizing radiation of the Of stars HD 228841 and HD 193595
belonging to Cyg~OB1. The chain of CGs extends radially with
respect to these stars. The morphology of CGs in the IR bands is
as a whole consistent with the optical images: dust head and tail
bordered with bright IR and/or optical rims. The optical emission
encircles the bright IR globules.

The ambient ionized gas is characterized by high-velocity features
in the H$\alpha$ line (up to 60--80~km~s$^{-1}$ with respect to
the systematic velocity); this is a consequence of the action of
the wind from these stars. The gas swept out by the wind
circumflows the family of cometary globules, since high velocities
are observed along the entire chain of CGs. We have found no
obvious kinematic signs of braking of the high-velocity gas at the
interaction with the `Three-headed globule'. Indirect indications
to the interaction of the wind from the star and the family of CGs
can be the presence of a CO cavern surrounding the eastern
globules and fine IR filaments curled away from the stars.

We have observed a compact stellar object in the southern head of
the brightest globule -- the star HBHA~3703-01 identified with the
infrared source IRAS~20153+3850. Our multicolour photometric
observations allow us to estimate the spectral type of HBHA
3703-01 as B(5--6)V and to measure its colour excess $E(B-V) =
1.18$~mag. Low-resolution spectral observations of HBHA~3703-01
have revealed the broad H$\alpha$ emission with $\textrm{FWHM} =
3.7\pm 0.5$~\AA{}, which corresponds to a high velocity, $\Delta
v=169\pm 25$~km~s$^{-1}$. The absolute visual magnitude of the
star obtained under the assumption that the star belongs to the
Cyg~OB1 starburst complex locates it on the H-R diagram near ZAMS
in the region of Herbig Ae/Be objects.

HBHA~3703-01 is surrounded by the reflection nebula GM~2-39; its
diameter is about $30^{\prime\prime}$. The nebula is distinguished
against the background of the overall molecular cloud just because
of the presence of the star, which illuminates a part of the
interstellar medium nearest to the star and results in the
formation of an optical reflection nebula.

Using archival data of observations of HBHA 3703-01 obtained by
the \emph{Spitzer} Space Telescope, we have estimated the fluxes
in the 3.6, 4.5, 5.6, 8 and 24~$\mu$m bands. (On all IR images
HBHA~3703-01 looks as a starlike object surrounded with a nebula;
the emission of this nebula is subtracted from the measured
fluxes.) We have also used our photometric data for HBHA 3703-01
together with the 2MASS magnitudes ($J$, $H$, $K_{s}$). According
to these data, we have derived the SED of HBHA~3703-01 in the
0.44--24~$\mu$m range. All observations were corrected for the
interstellar extinction with $E(B-V) = 0.87$~mag.

The obtained SED can be rather well described within the
framework of the model of emission of the central source
(a blackbody with temperature $T = 15,000$~K) surrounded
by a dust envelope with parameters listed above.

The derived dust density distribution in the
envelope $n(r)\propto 1/r^{1.5}$ is typical of the
envelopes of Herbig Ae/Be stars.

\section*{Acknowledgments}
This work was based on the data of IFP observations in the
H$\alpha$ line, spectral observations on the 125-cm reflector and
photometric observations on the 60-cm reflector of the Crimean
Laboratory of the Sternberg Astronomical Institute, Moscow State
University. We used archival data of the \emph{Spitzer}, AKARI and
2MASS observations, radio data in the CO line and the database of
the Centre des donn\'ees astronomiques, Strasbourg, France. This
research was supported by the Russian Foundation for Basic
Research (project codes 10-02-00091-a, 12-02-31356). OVE is also grateful for
the financial support of the `Dynasty' Foundation. We thank
S.V.~Antipin who made some of the photometric observations. We thank the anonymous referee for the constructive comments that improved the quality of the paper.

\bsp

\label{lastpage}

\end{document}